\documentclass[10pt]{article}
\usepackage[OE]{express}

\newcommand{\p}{\partial}
\newcommand{\ep}{\varepsilon}

\newcommand{\om}{\omega}

\newcommand{\ta}{\theta}

\newcommand{\be}{\begin{equation}}
\newcommand{\ee}{\end{equation}}
\newcommand{\ba}{\begin{eqnarray}}
\newcommand{\ea}{\end{eqnarray}}
\usepackage{comment}
\begin{document}
	\title{Soliton and quasi-soliton frequency combs due to second harmonic generation in microresonators}
	
\author{
Alberto~Villois,\authormark{1,*} 
and Dmitry V. Skryabin~\authormark{1,2}}
	
	\address{
		\authormark{1} Department of Physics, University of Bath, Bath BA2 7AY, UK \\
			\authormark{2} Russian Quantum Centre, Skolkovo 143025, Russia\\
		  \authormark{*} \textcolor{blue}{a.villois@bath.ac.uk}}
	



\begin{abstract}
We report how a doublet of the symmetric oppositely tilted bistable resonance peaks in a microring resonator with quadratic nonlinearity set for generation of the second harmonic can transform into a Kerr-like peak on one side of the linear cavity resonance and into a closed loop structure disconnected from the quasi-linear resonance on the other. Both types of the nonlinear resonances are associated with the formation of the soliton combs for dispersion profiles of a typical LiNbO$_3$ microring. We report bright quasi-solitons propagating on a weakly modulated low intensity background when the group velocity dispersions have the opposite signs for the fundamental and second harmonic. We also show exponentially localized solitons when the dispersion signs are the same. Finally, we demonstrate that the transition between these two types of soliton states is associated with the closure of the forbidden gap in the spectrum of quasi-linear waves.
\end{abstract}

\section{Introduction}
Frequency comb generation in microring resonators with Kerr nonlinearity has been intensely studied in the last decade. These studies have demonstrated a plethora of novel solitonic effects with immediate applications in the precision measurements and optical signal processing \cite{kip1,str}. Using nonlinear effects other than Kerr nonlinearity for comb generation is an active research area. One can mention here examples of the comb generation due to cascaded Raman and Brillouin effects \cite{ram1,ram2,bril1,bril2}, polariton  \cite{pol1,pol2} and
graphene nonlinearities \cite{graph}, and  quadratic, $\chi^{(2)}$, nonlinear effects \cite{q1,q2,q3,q5,q6,q8,q10,q11,q12,ibm,q13,q14,lon}. The latter hold a particular promise. $\chi^{(2)}$ materials are widely available and well studied for frequency conversion and soliton applications. They provide high levels of nonlinear response allowing to reduce pump power requirements. Importantly, phase matching for the three wave mixing in $\chi^{(2)}$ materials and the soliton formation conditions do not require
anomalous dispersion, as it takes place in Kerr, $\chi^{(3)}$, materials \cite{rev1}. This significantly extends the playing field for the comb research in terms of enhancing flexibility of the pump source choices and generating combs over wider spectral ranges between mid-infrared and ultraviolet.

Following pioneering observations of 2nd harmonic generation in a whispering gallery mode resonator~\cite{n1}, research into frequency conversion for those devices was limited by the difficulty in achieving a wide mode spectra generated in the proximity of both pump and signal waves over some period of time, see~\cite{n2} for a review. Only recently the first convincing experimental observation of a broad frequency combs and a cascaded multi-wave mixing has been reported in a micro-resonator with $\chi^{(2)}$ nonlinearity \cite{q12,ibm}. Since most practical low noise and, at the same time, broad combs are provided by bright dissipative optical solitons \cite{kip1} in microresonators, theoretical understanding of cavity solitons supported by quadratic nonlinearity is essential. We note that first theories on $\chi^{(2)}$ cavity solitons date two decades back, we recall for example early studies on diffractive solitons in planar microcavities, see, e.g., \cite{led1,me1}. However, $\chi^{(2)}$ cavity solitons had not received much of the experimental attention so far, in contrast to their Kerr couterparts \cite{kip1}. Nowadays, due to the growing interest in comb generation in  $\chi^{(2)}$ microresonators \cite{q1,q2,q3,q5,q6,q8,q10,q11,q12,ibm,q13,q14,lon}, the theoretical studies of quadratic solitons are particularly relevant to  stimulate and guide growing experimental efforts in this area. We recall, for example, the recent experiment in a bow-tie cavity, on the observation of $\chi^{(2)}$ simultons (coupled bright and dark pulses) under conditions where the 2nd order dispersion is negligible~\cite{q10}. The theoretical problems that are relevant at present and that are connected to the study below are the frequency comb generation under conditions of the normal and mixed group velocity dispersion signs at the fundamental and second harmonics and dispersive wave emission by solitons, which is expected to have a different effect in a ring geometry compare to the waveguides.

The paper is structured as follows: Section 2 introduces the reader into the Lugiato-Lefever model and maps its parameters on the dispersion profiles 
corresponding to a LiNbO$_3$ microring wire resonator. Section 3 covers single mode solutions and their nonlinear resonances. In Section 4 theoretical considerations concerning the criteria for the existence of quadratic solitons are discussed. In particular, we  explain why these type of solutions belong to a class of gap solitons. Numerical results on the soliton and quasi-soliton states are described in Section 5. 
Our work has been performed and results have been obtained independently from   \cite{q13} on comb solitons due to second harmonic generation published in December 2018. While finalising our material for publication we have not included data overlapping with  \cite{q13} related to the case of the symmetric resonance splitting as in our Fig. 2(a).

\begin{figure}[t]
	\centering
	\includegraphics[width=\textwidth]{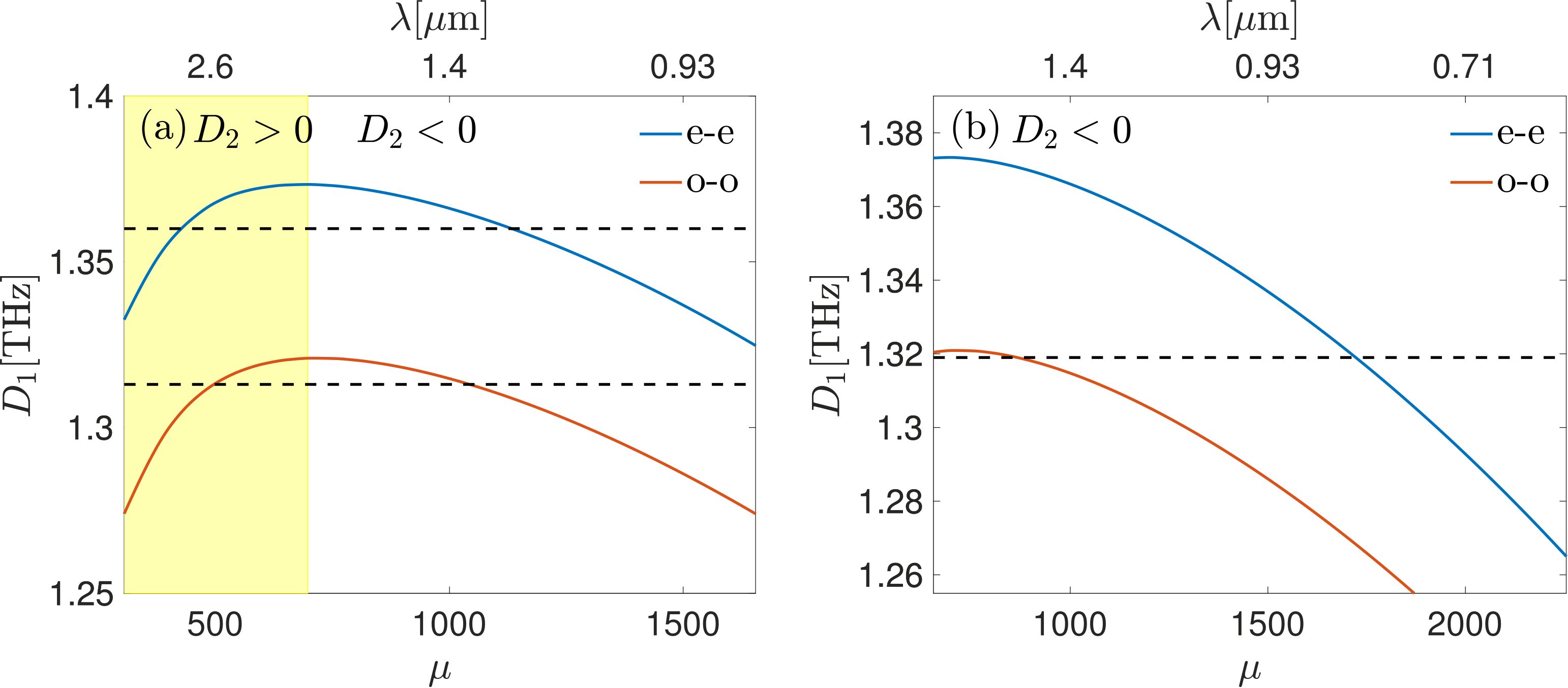}
	\caption{(a) $D_1$  vs modal number $\mu$ (bottom axis) and wavelength $\lambda$ (top axis). Dashed lines mark points of $D_1$ matching between the fundamental and second harmonics, such that $D_{1f}=D_{1s}$. (a) Fundamental and 2nd harmonics are either both ordinary or both extraordinary waves.  $D_{1f}=D_{1s}=1.31\,$THz, $D_{2f}=88.2\,$MHz, $D_{2s}=-42.1\,$MHz at the matching points for the red line and $D_{1f}=D_{1s}=1.36\,$THz, $D_{2f}=94.3\,$MHz, $D_{2s}=-40.2\,$MHz for the blue line. (b) Fundamental is ordinary polarized and 2nd harmonic is the extraordinary polarized. Here $D_{1f}=D_{1s}=1.32$THz, 
	$D_{2f}=-24.1$MHz, $D_{2s}=-86.9$MHz at the matching points. \label{fig1}}
\end{figure}

\section{Lugiato-Lefever model for a microring with $\chi^{(2)}$ nonlinearity}
To simulate the spatio-temporal evolution of the fundamental and 2nd harmonic fields in a microring resonator we use the following model equations
\ba\label{e1}
&& i\partial_T\Psi_f=\left(\omega_{f}-iD_{1f}\partial_{\phi}-\tfrac{1}{2}D_{2f}\partial^2_{\phi}\right)\Psi_f-i\kappa_f\Psi_f-\alpha \Psi_f^*\Psi_s -\alpha h e^{-i\omega_p T},\\
&& i\partial_T\Psi_s=\left(\omega_{s}-iD_{1s}\partial_{\phi}-\tfrac{1}{2}D_{2s}\partial^2_{\phi}\right)\Psi_s-i\kappa_s\Psi_s-\alpha\Psi_f^2.
\label{e2}\ea 
Here $T$ is the physical time and $\phi$ is the polar angle varying along the resonator circumference. The amplitudes of the dimensionless electric fields are given by $\Psi_{f}$ and $\Psi_{s}$ for the fundamental and for the 2nd harmonic, respectively. $\omega_p$ is the frequency of the CW pump laser while $\omega_f$ corresponds to its closest resonant frequency in the linear spectrum of the microring. Similarly, $\omega_s$ is the nearest resonance to the 2nd harmonic of the pump, $2\omega_p$.  We have assumed the exact angular phase matching conditions here, that can be achieved through the quasi-phase matching technique \cite{n2}. Accounting for the time derivatives and the bracketed expressions in both Eqs. (\ref{e1}) and (\ref{e2}) allows to approximate the microring dispersion around the pump and its 2nd harmonic. The coefficients $D_{1f,1s}$ correspond to the local free spectral ranges (FSRs) of the microresonator and $D_{2f,2s}$ set the FSR  dispersion (group velocity dispersion). Finally, $\kappa_f$ and $\kappa_{s}$ represent the photon loss rates, $h$ is the dimensionless amplitude of the CW pump and $\alpha$ is a scaling parameter with units of frequency.

\begin{figure}[t]
	\centering
	\includegraphics[width=\textwidth]{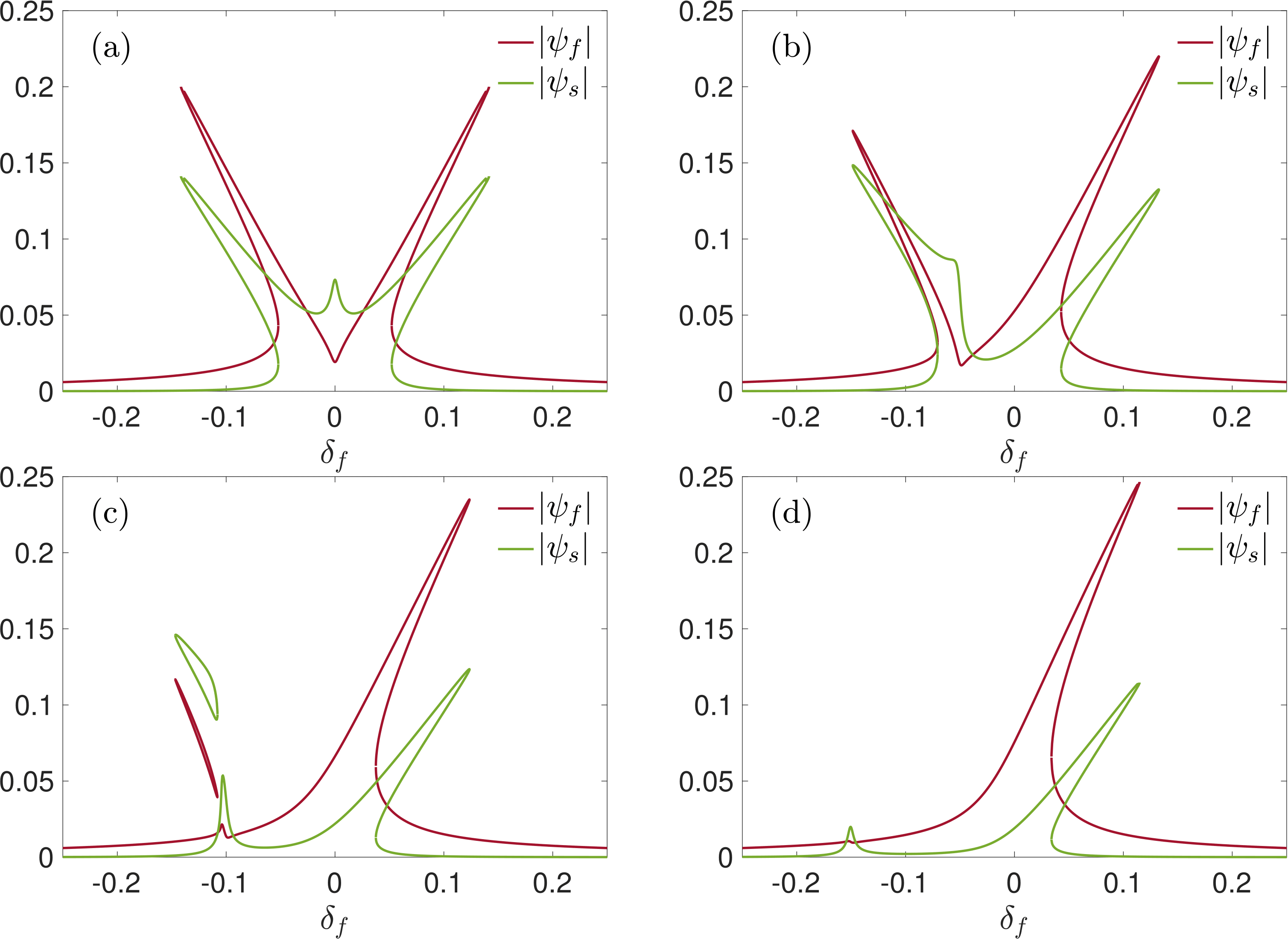}
	\caption{Homogeneous solutions for both fundamental (red lines) and second harmonic (green lines) fields vs detuning $\delta_f$. Here $h=0.0015$ and $\gamma_f=\gamma_s=0.005$. $\delta_s=2\delta_f+\ep$ and  $\epsilon=0$ (a), $\epsilon=0.1$ (b),  $\epsilon=0.2$ (c),  $\epsilon=0.3$ (d). \label{fig2}}
\end{figure}

Our primary interest here is in finding the soliton combs, therefore we 
look for shape preserving pulses rotating around the microring with the constant frequency $\Omega$: $\Psi_{f}(\phi,T)=\psi_{f}(\theta,t)e^{-i\om_pT}$, $\Psi_{s}(\phi,T)=\psi_{s}(\theta,t)e^{-i2\om_pT}$, where $\phi=\theta-\Omega T$ and $t=T/T_f$ is the dimensionless time normalized to the pulse round trip time $T_f=2\pi/D_{1f}$ at the fundamental frequency $2 \pi D_{1f}$. Then $\alpha$ is conveniently fixed as $\alpha=1/T_f$. Let us stress that $\Omega$ is the same for the fundamental and 2nd harmonic fields, so that if $\psi_{s,f}$ are assumed to be $t$ independent, then both fields will be permanently locked one to another. We also note that this assumption can be practically satisfied only if $D_{1f}$ and $D_{1s}$ are sufficiently close, as will be discussed in detail below. It is therefore convenient to assume that
$\Omega=D_{1f}(1-V)$, where $V$ is proportional to the FSR shift induced by the nonlinearity. The above substitutions result in the dimensionless system of equations convenient for our numerical and theoretical studies:
\ba\label{e3}
&& i\p_t\psi_f=-2\pi\left(iV\p_{\theta}+\frac{1}{2}d_{2f}\p^2_{\theta}\right)\psi_f+(\delta_f-i\gamma_f)\psi_f-\psi_f^*\psi_s -h, \\
&& \label{e4}
i\p_t\psi_s=-2\pi\left(i(V-U)\p_{\theta}+\frac{1}{2}d_{2s}\p^2_{\theta}\right)\psi_s+(\delta_s-i\gamma_s)\psi_s- \psi_f^2.
\ea
Here $\gamma_{f,s}=2\pi\kappa_{f,s}/D_{1f}$ and $d_{2f,2s}=D_{2f,2s}/D_{1f}$ are the normalized losses and dispersion parameters. We also introduce $\delta_f=2\pi(\omega_f-\omega_p)/D_{1f}$ and $\delta_s=2\pi(\omega_s-2\omega_{p})/D_{1f}=2\delta_f+\epsilon$ which represent the normalized detuning parameters. Due to the discrete nature of the spectrum of a microresonator and dispersion, even if we tune $\omega_p$ to perfectly match $\omega_f$, the 2nd harmonic $\omega_s$ will in general be off-resonance. To carefully account for that we define $\epsilon=2\pi(\om_s-2\om_f)/D_{1f}$ to be a constant offset between the resonance frequency $\om_s$ nearest to $2\omega_p$ and the doubled frequency $\om_p$ of the fundamental resonance. The importance of $\ep$ is in shaping the nonlinear response of the
microresonator, and its effect will be discussed in detail in the following section. Finally, $U=(D_{1f}-D_{1s})/D_{1f}$  is the normalized difference between the FSRs at the fundamental and at 2nd harmonic frequencies. 
This parameter characterizes the difference of the pulse rotation frequencies of the fundamental and 2nd harmonic fields. In the soliton regime $U$ is compensated through nonlinear effects \cite{rev1}, so that the two fields lock and propagate together. $U$ can also be minimised or eliminated by adjusting the pump frequency and cavity dispersion, see below. We note, that in the later case $U=0$ implies the selection of a particular value of $\epsilon$.

Considering a toroidal microring with the large radius $100\,\mu$m and the small one $1\,\mu$m made of LiNbO$_3$  and using the theory in \cite{gorod} we estimate values of the dispersion and detuning parameters to be used in our numerical analysis below. These estimates can also be used as a qualitative guide for the planar on-chip wire microresonators, see, e.g., \cite{q3,lon}.
Considering the same polarisation states for the fundamental and 2nd harmonic fields we found that it is possible to match FSRs at both frequencies across the zero dispersion point, see Fig. 1(a). Using an ordinary wave for the fundamental and the extraordinary for the 2nd harmonic it is possible to match $D_{1f}$ to $D_{1s}$ maintaining $D_{2f}$ and $D_{2s}$ negative, see Fig. 1(b).
Note, that the negative/positive $D_2$ corresponds here to the normal/anomalous dispersion.

\section{Nonlinear resonances and bistability}
We start our analysis of the nonlinear microring states from the CW-solutions that correspond to homogeneous solutions of Eqs. (\ref{e3}) and (\ref{e4}) ($\p_t=\p_\ta=0$) and we check how quadratic nonlinearity distorts the shape of the linear resonance response.
Figure 2 shows how the amplitude of CW solutions vary with $\delta_f$ for several values of $\ep$. Note that changing $\delta_f$ corresponds experimentally to tuning the pump laser frequency $\om_p$ around $\omega_f$. We also stress that during this process the value of $\ep$ remains constant.
In Fig. 2(a) we consider the ideal case when $\ep = 0$. For such a value of $\ep$ the linear resonance splits symmetrically and two tilted bistable resonances are formed ~\cite{q13}. To capture all relevant properties of the system one needs to account for $\ep\neq0$, which leads to a shift and reshaping of the structure of nonlinear resonances. When $\ep>0$, see Figs. 2(b)-2(d), the most of reshaping appears in the negative range of values of $\delta_f$, while the situation is reversed when $\epsilon<0$. By further increasing the value of $\epsilon$, the negatively tilted resonance undergoes a distortion which narrows down its bistability range. This process ultimately leads to the detachment of the tilted resonance from the CW low amplitude background, see Fig. 2(c). We stress that the latter scenario can have an important implication concerning the experimental technique used to achieve solitons in microresonators. It is well known that the Kerr soliton combs can be obtained by adiabatic scanning of the laser pump across a resonance~\cite{kip1}. In this way, unstable high amplitude CW solutions tend to decay into nearby soliton states. However, the presence of a  resonance loop disconnected from low amplitude state may compromise this experimental technique, which suggests that hard excitation methods might be a more suitable choice to achieve comb-soliton states.     While stability properties of the upper branches of the CW-solutions involve many details reporting on which goes beyond our present scope, the low amplitude solution are linearly stable through ranges of the soliton existence shown with full lines in Fig. 5 below.
\begin{figure}[t]
	\centering
	\includegraphics[width=\textwidth]{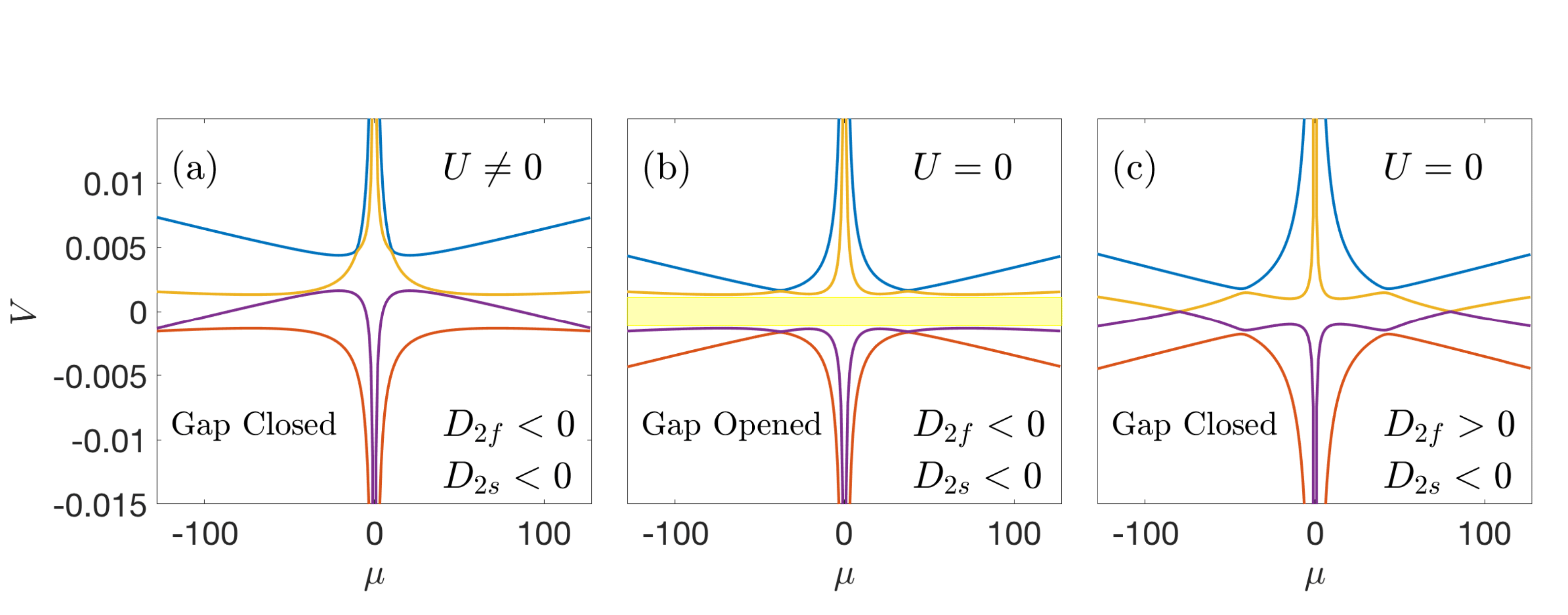}
	\caption{
	Plots of the 4 roots of $P(V)$. Each root is shown with a different color.
	(a) The $V$-gap is closed due to group velocity mismatch effects, i.e. $U\ne 0$: $U=-0.003$, $d_{2f}=-1.83 \times10^{-5}$, $d_{2s}=-6.59\times10^{-5}$, $h=0.0015$, $\delta=-0.3$, $\epsilon=0.51$. (b) The $V$-gap is opened in the case of GVDs being negative at both frequencies: $U=0$, $d_{2f}=-1.83 \times10^{-5}$, $d_{2s}=-6.59\times10^{-5}$, $h=0.0015$, $\delta=-0.3$, $\epsilon=0.51$. (c) The $V$-gap is closed in the case of group velocity dispersions with the opposite signs: $U=0$, $d_{2f}=6.89 \times10^{-5}$, $d_{2s}=-2.94\times10^{-5}$, $h=0.0015$, $\delta=0.05$, $\epsilon=0.49$.\label{fig3}}
\end{figure}

\section{Soliton existence conditions}
In this Section we analyse the comb soliton existence conditions with two different approaches.
For sufficiently large $\ep$, see Fig. 2(d), the nonlinear resonance effects for negative detunings become suppressed. The system enters into the Kerr-like regime, also called the cascading nonlinearity regime~\cite{rev1}. Here the 2nd harmonic simply adiabatically follows the fundamental one, such that $\psi_s\simeq \psi_f^2/\delta_s$. The latter approximation reduces Eqs. (\ref{e3}) and (\ref{e4}) to the standard Lugiato-Lefever model for stationary comb solitons \cite{kip1}: $\pi d_{2f}\p^2_\ta\psi_f\simeq \delta_f\psi_f-|\psi_f|^2\psi_f/\delta_s -(h+2i\pi V\p_\ta+i\gamma_f)\psi_f$. One of the advantages of working with a cascading $\chi^{(2)}$ nonlinearity, over a pure Kerr system, is that the sign of the cascading nonlinearity is controlled
by the value of $\delta_s=2\delta_f+\ep$. Hence bright soliton combs can be expected for both normal ($D_{2f}<0$, $\delta_{f,s}<0$) and anomalous ($D_{2f}>0$, $\delta_{f,s}>0$) dispersion. 

\begin{figure}[t]
\centering
\includegraphics[width=\textwidth]{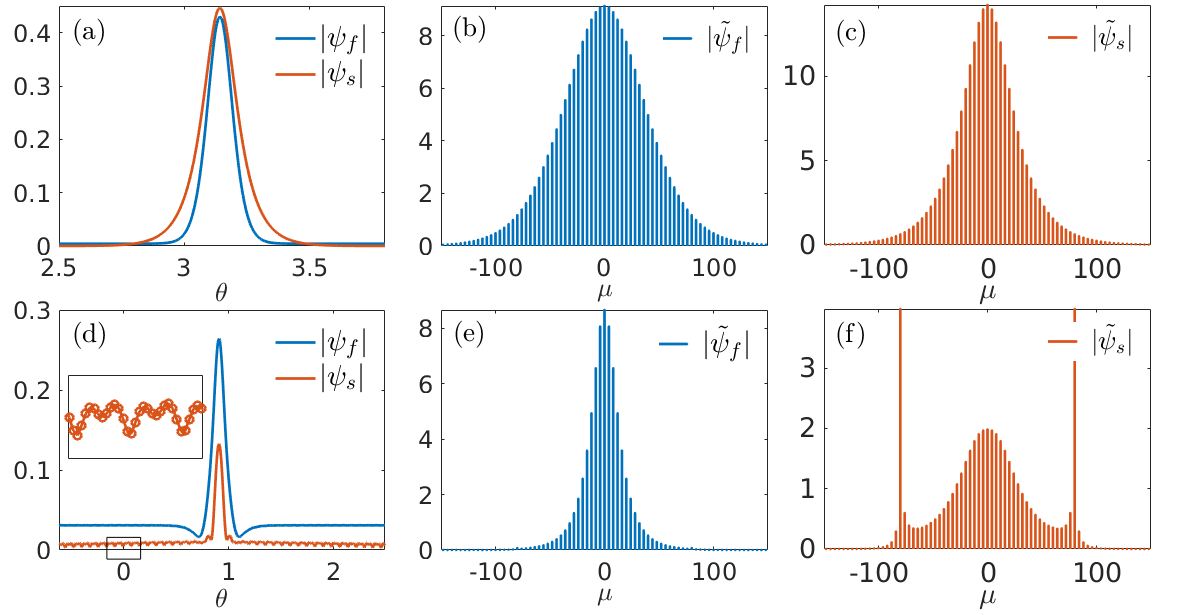}
\caption{Soliton profiles and the corresponding frequency comb spectra. Red/blue is the fundamental/second harmonic fields. Top row is the case of the comb solitons with the exponentially damped tails inside the gap as in Fig. 2(b): $d_{2f}=-1.83 \times10^{-5}$, $d_{2s}=-6.59\times10^{-5}$, $h=0.0015$, $\delta=-0.3$, $\epsilon=0.51$, $\gamma_f=\gamma_s=0.0012$.
Bottom row is the case of the comb solitons with the oscillatory tails, when the gap as in Fig. 2(c): $d_{2f}=6.89 \times10^{-5}$, $d_{2s}=-2.94\times10^{-5}$, $h=0.0015$, $\delta=0.05$, $\epsilon=0.49$ and $\gamma_f=\gamma_s=0.005$. The inset corresponds to highlighted rectangular area.\label{fig4}}
\end{figure}

Since the role played by the 2nd harmonic field in the soliton existence condition remains obscure in the cascading limit analysis, we are analysing the soliton existence conditions using a more formal, but physically grounded approach.
For high quality microresonators the loss term can be disregarded in first approximation. This approximation allows to analyse qualitatively and quantitatively the role played by the velocity parameter $V$
and by the 2nd harmonic field in the soliton formation. The low amplitude state of the nonlinear resonances is known to serve as the background state for the bright comb solitons. Thus one can say that the necessary condition for the soliton existence is that this state does not support the extended modes representing quasi-linear small amplitude perturbations distorting the soliton. In particular, we are anticipating that for some parameter values there should exists a gap in $V$-values such that for all modal numbers $\mu$ one can not find a $V$ value inside the gap giving a physically realizable quasi-linear periodic wave forms. We assume that such quasi-linear waves can be sought as $t$-independent solutions of Eqs. (\ref{e3}) and (\ref{e4}) in the form
\be\label{e5}
\psi_{f,s}(\ta)=\psi_{f0,s0}+\zeta_{f,s} e^{i \mu \theta}+\xi_{f,s}^* e^{-i \mu \theta}.
\ee
Here $\psi_{f0}$ and $\psi_{s0}$ are the fundamental and 2nd harmonic  amplitudes of the lowest amplitude state within the bistability ranges of the nonlinear resonances, while $\zeta_{f,s}$ and $\xi_{f,s}$ are small amplitudes of the sought waves.
We note that $\mu$ is one of the modes of the microresonator counted from zero. Here $\mu=0$ corresponds to the mode associated to $\om_f$. Substituting Eqs. \eqref{e5} into Eqs. \eqref{e3} and \eqref{e4} and linearising for small $\zeta_{f,s}$ and  $\xi_{f,s}$ we find that, for the solutions to exist, the velocity parameter has to satisfy the fourth order equation $a_4V^4+a_3V^3 +a_2V^2+a_1V+a_0=0$, where all $a_n$'s are parameterized by $\mu$. 

Figure 3 shows typical plots of four roots of $V$ vs $\mu$.  Figure 3(a) demonstrates the situation when dispersion signs are the same and the mismatch of group velocities $U$ is sufficiently large to overcome the nonlinear effects
to pull two pulses apart. This is the physical argument prohibiting the existence of the mutually locked fundamental and 2nd harmonic pulses. This argument matches with our gap analysis results showing that  the small amplitude periodic waves are allowed for all $V$'s for these system parameters. However, when $U$ is sufficiently small or zero, then an interval (gap) of $V$ values, where the small amplitude periodic waves are forbidden, is emerging, see Fig. 3(b) and cf. Fig. 1(b). We expect that the bright comb solitons can exist under these conditions since a pulsed excitation with, e.g., $V=0$, will not couple to a continuum of quasi-linear waves. When $D_{2f}>0$ and $D_{2s}<0$, see Fig. 3(c) and cf. Fig. 1(a), the gap between the two roots closes for some relatively large $\mu=\mu_{cr}$ at $V=0$ and even when $U=0$ is chosen. This implies, in the spirit of the theory of dispersive wave emission by solitons \cite{rmp},  that the quasi-soliton excitation with $V=0$ is going to develop tails with a small amplitude dispersive wave at $\mu=\mu_{cr}$. If $\mu_{cr}$ is sufficiently large ($\mu_{cr}$ tends to infinity with $D_{2f}\to 0$), then an impact of the dispersive waves on the soliton core is expected to be negligible, see previous theories for the dispersive wave emission by Kerr solitons \cite{rmp}.

\begin{figure}[t]
\centering
\includegraphics[width=\textwidth]{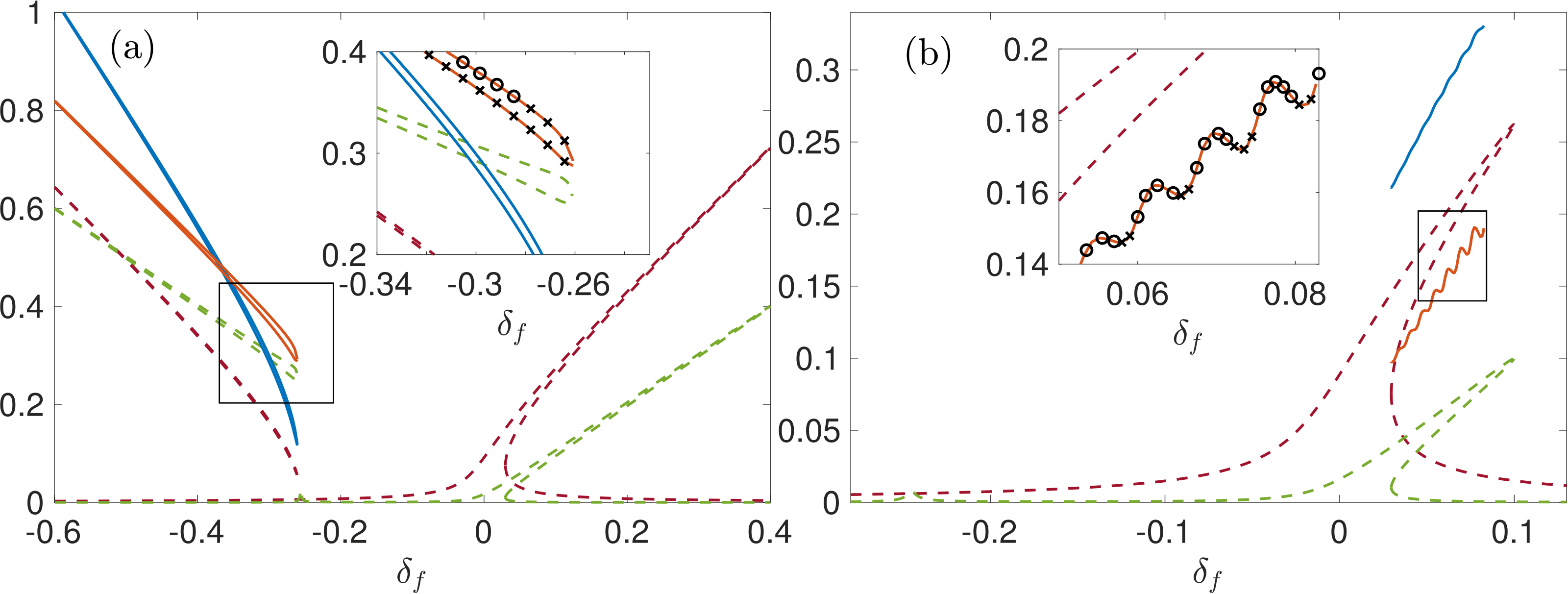}
\caption{Maximal amplitude of the comb soliton pulses (full lines: blue/red is fundamental/2nd harmonic) and CW-solutions (dashed lines: magenta/green is fundamental/2nd harmonic). (a) is the case of the same dispersion signs and relatively small $\gamma_{f,s}$: $U=0$, $d_{2f}=-1.83 \times10^{-5}$, $d_{2s}=-6.59\times10^{-5}$, $h=0.0015$, $\epsilon=0.51$, $\gamma_{f,s}=0.0012$. The inset corresponds to highlighted rectangular area. Circles/crosses mark linearly stable/unstable solitons. (b) is the case of different GDV signs and of the larger $\gamma_{f,s}$ (which suppress the negatively tilted resonance): $U=0$, $d_{2f}=6.89 \times10^{-5}$, $d_{2s}=-2.94\times10^{-5}$, $h=0.0015$, $\epsilon=0.49$, $\gamma_{f,s}=0.005$. The inset corresponds to highlighted rectangular area. Circles/crosses mark linearly stable/unstable quasi-solitons.\label{fig5}}
\end{figure}

\section{Bright  soliton combs}
We will now numerically investigate existence and dynamics of bright soliton solutions, which spectrally correspond to the frequency comb states due to 2nd harmonic generation. Throughout this study, the parameters $D_{2f,2s}$ and $\ep$ are chosen using the dispersion data in Fig. 1. Numerical simulations to find soliton families in the time-independent version of Eqs. \ref{e3} and \ref{e4} are performed using a Newton method in combination with a Bi-conjugate gradient method implemented in the Fourier space, such that periodic boundary conditions are accounted for automatically. To simulate soliton dynamics we use standard pseudo spectral approach. All our results are obtained considering 2048 modes.

\begin{figure}[t]
\centering
\includegraphics[width=1.\textwidth]{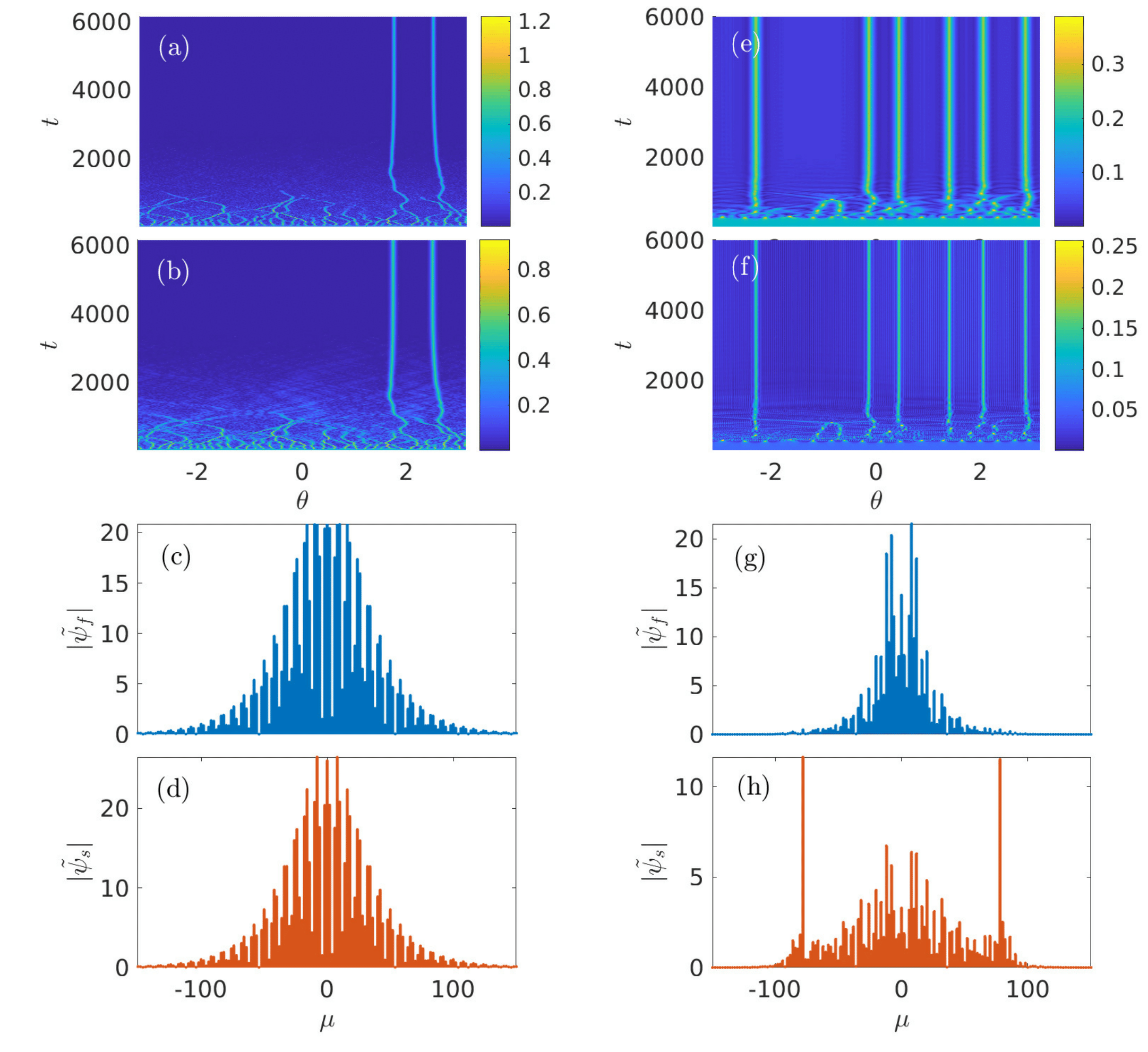}
\caption{Numerical simulations of Eqs. (3) and (4) using as initial condition the CW corresponding to the upper branches of the tilted resonance. After early time dynamics we show generation of comb solitons (left column) and quasi-solitons (right column). Spectra after $6000$ cavity round trips are shown.  Left/right column parameters are the same as the ones used in Fig. 4 top/bottom row. \label{fig6}}
\end{figure}

We start from the case when dispersions are normal ($D_{2f,2s}<0$) for both fundamental and 2nd harmonic, which is the case shown in Fig. 1(b). Bright soliton combs in a material with positive Kerr do not exist under these common conditions. However, $\chi^{(2)}$ effects allow their existence.  The dimensionless parameters in Eqs. (\ref{e3}) and  (\ref{e4}) which allow for the FSR matching are estimated to be $\epsilon=0.51$,  $d_{2f}=-1.83\times 10^{-5}$ and $d_{2s}=-6.59\times 10^{-5}$. We note that these 
parameters are the same as the ones used in Fig. 3(b). This suggests the presence of an open gap in the soliton velocity parameter $V$. More specifically, both the $V$-gap analysis and the cascading approximation predict the existence of solitons when negative values of detuning are considered. Solitons are indeed found when $\delta_f<0$. Figures 4(a)-4(c) show the spatial profiles of the two soliton components and their corresponding frequency combs. The structure of both nonlinear resonances and soliton families for different values of $\delta_f$ are shown in Fig. 5(a). We note that, continuing the soliton family solutions in $\delta_f$, we found a closed loop structure similar to the one for the cw state, see the zoomed inset in Fig. 5(a). This is true for both fundamental and 2nd harmonic solitons family. Performing linear stability analysis we have confirmed  that the low amplitude soliton state in the loop is unstable. Its dynamics, in fact, leads either to the excitation of an attractor in the proximity of the high amplitude state or to the decay into the lower amplitude stable background. On the other hand, the soliton state with the larger amplitude can be stable. We note the presence of some instabilities for small values of $\delta_f$, see Fig. 5(a), but reporting and studying peculiarities of those goes beyond the aim of this work.

We will now discuss the case of different signs of dispersions: anomalous ($D_{2f}>0$) for the fundamental harmonic and normal ($D_{2s}<0$) for the 2nd harmonic. In particular, we will consider the scenario when both fields are extraordinary polarized, see Fig. 1(a). The dimensionless parameters in Eqs. (\ref{e3}) and (\ref{e4}) which allow for the FSR matching are given by $\epsilon=0.49$,  $d_{2f}=6.89\times 10^{-5}$ and $d_{2s}=-2.94\times 10^{-5}$. Similarly to the previous case, these parameters have already been used in Fig. 3(c), which does not show any open $V$-gap. Despite the absence of an open gap, the Lugiato-Lefever model with Kerr type of nonlinearity, is  approximately valid in the cascading limit, and still predicts the existence of bright solitons for positive detunings. Solitons are indeed found when $\delta_f>0$ and their  profiles and the corresponding spectral combs are shown in Figs. 4(d)-4(f). In order to explain this contradicting result one needs to account for the finite size of the microresonator. The theory of dispersive waves emission by solitons \cite{rmp} implies that solitons in this situation are not truly localized objects (quasi-solitons), but they will transfer energy to small amplitude dispersive waves with the modal index closely matching  the mode number corresponding to the crossing of different $V$-branches within the former gap. In our case the critical mode number is given by $\mu_{cr}=\pm 82 $. However, being in a confined system the transfer of energy from the soliton to the dispersive (Cherenkov) radiation does not necessarily causes the decay of the soliton. This is due to the fact that the radiation travels along the ring and transfers energy back to the soliton. A family of these quasi-solitons was numerically traced in $\delta_f$ and  is shown in Fig. 5(b) together with the corresponding CW solutions. We note that here the absence of the left tilted resonance is due to the relatively large choice of $\gamma_{f,s}=0.005$.

A feature of the quasi-solitons in a ring geometry is that the radiation creates an effective potential landscape for the soliton which makes it more robust with respect to the translational shifts. This has been reported for Kerr comb solitons with the 4th order dispersion \cite{kar} and is expected here. The radiation amplitude is significantly stronger for the second harmonic, which corroborates the hypothesis that the gap closure in Fig. 3(c) involves the dispersion branches $V(\mu)$ associated with the 2nd harmonic field. In fact, if the cascading limit is considered, the dispersive branches for the 2nd harmonic are eliminated and exponentially damped tails are recovered. We stress that these strong dispersive resonances substantially amplify spectral tails of the comb around the 2nd harmonic, which can be utilised for practical purposes. 

Finally,  we confirm the comb generation using dynamical simulations. More specifically, we found that the solitons emerge from modulational instabilities of the upper branch of the CW solutions, see Figs. 6(a)-6(b) and Fig. 6(e)-(f). Using the same parameters of Fig. 5, plots in Figs. 6(a)-6(d) represent the case with the same signs of dispersions, while plots in Figs. 
6(e)-6(h) correspond to the opposite signs. These simulations not only reproduce combs similar to the ones in Fig. 4, but also confirm the robustness of $\chi^{(2)}$ comb solitons. We note, that solitons also remain stable against perturbations coming from the background radiation waves, which are particularly strong during the initial stages of the soliton evolution.

\section{Conclusions}
We have developed a theory that demonstrates that comb solitons in microring resonators with quadratic nonlinearity belong to a class of gap-solitons. We demonstrated the existence of the families of comb solitons for normal dispersion at both the fundamental and second harmonic fields. Some of these solitons exist under conditions of the strongly modified nonlinear resonances, that are not smoothly connected to the linear resonance tails. 
When the solitonic gap is closed through the action of the opposite dispersion signs at the fundamental and second harmonic, we have found that the spectral tails of the second harmonic comb acquire strong resonance peaks corresponding to the dispersive wave emitted by the soliton. The associated spectral broadening may find practical applications. Dispersion values and group velocity matching conditions in our work have been calculated for the resonator made of the $1\mu$m wide LiNbO$_3$ waveguide.

\section*{Funding}
The Leverhulme Trust (RPG-2015-456); H2020 (691011, Soliring); Russian Science Foundation (17-12-01413).

\section*{Acknowledgments}
We thank I.~Breunig and J.~Szabados for helpful suggestions and discussions.
	
\end{document}